\documentstyle[12pt]{article}
\newcommand{\mg}{m_{\tilde{g}}}
\newcommand{\mq}{m_{\tilde{q}}}

\newcommand{\beq}{\begin{equation}}
\newcommand{\eeq}{\end{equation}}
\newcommand{\beqn}{\begin{eqnarray}}
\newcommand{\eeqn}{\end{eqnarray}}
\newcommand{\stackM}{\stackrel{\scriptstyle >}{{ }_{\sim}}}
\newcommand{\stackm}{\stackrel{\scriptstyle <}{{ }_{\sim}}}
\begin{document}
\thispagestyle{empty}
\def\pubnum{407}
\def\data{January, 1997}

\begin{flushright}
{\parbox{3.5cm}{
UAB-FT-\pubnum

\data

hep-ph/9701392
}}
\end{flushright}

\vspace{3cm}
\begin{center}
\begin{large}
\begin{bf}
THE $\tan\beta-M_{H^{\pm}}$ BOUND FROM
INCLUSIVE SEMI-TAUONIC $B$-DECAYS
IN THE MSSM
\\
\end{bf}
\end{large}
\vspace{1cm}
J.A. COARASA, Ricardo A. JIM\'ENEZ, Joan SOL\`A\\

\vspace{0.25cm}
Grup de F\'{\i}sica Te\`orica\\
and\\
Institut de F\'\i sica d'Altes Energies\\
\vspace{0.25cm}
Universitat Aut\`onoma de Barcelona\\
08193 Bellaterra (Barcelona), Catalonia, Spain\\
\end{center}
\vspace{0.3cm}
\hyphenation{super-symme-tric sig-ni-fi-cant-ly ge-ne-ral}
\hyphenation{com-pe-ti-ti-ve}
\hyphenation{mo-dels}
\begin{center}
{\bf ABSTRACT}
\end{center}
\begin{quotation}
\noindent
\hyphenation{ob-ser-va-bles}
\noindent
We compute the ${\cal O}(\alpha_s)$ SUSY-QCD corrections to
the $W$ and charged Higgs mediated inclusive semi-tauonic $B$-decay,
$\bar{B}\rightarrow \tau\,\bar{\nu}_{\tau}\,X$.
Combining the SUSY contribution with the SM result obtained from
the heavy quark effective field theory, plus ordinary QCD corrections,
we find that the allowed region in
the $(\tan\beta,M_{H^\pm})$-plane could be
significantly modified by the short-distance supersymmetric effects.
Since the sensitivity to the
SUSY parameters other than $\mu$ (the higgsino mixing mass)
is rather low, the
following effective bound emerges for $\mu<0$
at the $2\,\sigma$ level:
$\tan\beta\stackm 0.43\,\ (M_{H^\pm}/{\rm GeV})$.
Remarkably, for $\mu>0$ there could be no bound at all.
Finally, we provide a combined $(\tan\beta,M_{H^\pm})$ exclusion plot
using our $B$-meson results together with the recent
data from top quark decays.
\end{quotation}

\baselineskip=6.5mm  

\newpage

Low-energy meson phenomenology can be a serious competitor
to high energy physics
in the search for extensions of the Standard Model (SM) of the strong
and electroweak interactions,
such as general two-Higgs-doublet models ($2$HDM's) and
Supersymmetry (SUSY). The simplest and most popular realization of the
latter is the Minimal Supersymmetric Standard Model\,\cite{MSSM}.
At present, the potential manifestations of the MSSM
are object of a systematic investigation. In this respect, $B$-meson
physics has been doing
an excellent job. On the one hand, the restrictions placed by
radiative $B^0$ decays $\bar{B}^0\rightarrow X_s\,\gamma$
(i.e. $b\rightarrow s\,\gamma$)
on the global fit analyses\,\cite{WdeBoer} to indirect precision
electroweak data have played a fundamental role.
In the absence of SUSY,
$b\rightarrow s\,\gamma$ alone is able
to preclude general Type II $2$HDM's
involving charged  Higgs masses
$M_{H^\pm}\stackm 200\,GeV$\,\cite{Barger}, thus barring the possibility
of the non-standard top quark decay $t\rightarrow H^+\, b$.
In fact, it is known that charged Higgs
bosons of ${\cal O}(100)\,GeV$ interfere constructively with the
SM amplitude of $b\rightarrow s\,\gamma$ and render
a final value of $BR(b\rightarrow s\,\gamma)$ exceedingly high.
This situation
can be remedied in the MSSM -- see later on -- where there may be
a compensating contribution from
relatively light charginos
and stops which tend to cancel the Higgs effects\,\cite{Ng}.
Thus, in the MSSM, the top quark decay
$t\rightarrow H^+\,b$ may well be open and could be a clue to
``virtual SUSY''\,\cite{CGJJS}.

On the other hand, semileptonic $B$-meson decays
can also reveal themselves as an invaluable probe for new physics.
In the specific case of the inclusive semi-tauonic
$B$-meson decays, $B^-\rightarrow \tau^-\,\bar{\nu}_{\tau}\,X$,
one defines the following ratio of rates
\beq
R={\Gamma(B^-\rightarrow \tau^-\,\bar{\nu}_{\tau}\,X)\over
\Gamma (B^-\rightarrow l^-\,\bar{\nu}_{l}\,X)}\,,
\label{eq:ratio}
\eeq
where $l=e,\mu$ is a light lepton. The SM prediction of this ratio (see later
on)
is a bit lower than the average experimental measurements. The discrepancy is
not dramatic, but it can be used to
foster or, alternatively, to hamper particular extensions of
the SM and, therefore, to restrict
or even rule out certain non-SM domains of the extended parameter space
where this ``$R$ anomaly'' would aggravate.
For example, the observable (\ref{eq:ratio}) is sensitive to
two basic parameters of generic
(Type II) $2$HDM's, namely the ratio of VEV's,
$\tan\beta=v_2/v_1$, and
the (charged) Higgs mass, $M_H\equiv M_{H^\pm}$. As a consequence,
the following upper bound at $1\,\sigma$
(resp. $2\,\sigma$) is claimed in the literature
 \,\cite{QCDHaber}:
\beq
\tan\beta<0.49\ (0.52)\,\ (M_H/{\rm GeV})\,.
\label{eq:tanMH1}
\eeq

To derive this bound, use is made of previous LEP 1 data
on semi-tauonic $B$-decays\,\cite{ALEPHL3}
\beq
BR(B^-\rightarrow \tau^-\,\bar{\nu}_{\tau}\,X)=(2.69\pm 0.44)\%\,,
\label{eq:input1}
\eeq
as well as of the former world average on semi-leptonic
$B$-decays\,\cite{world}
\beq
BR(B^-\rightarrow l^-\,\bar{\nu}_{l}\,X)=(10.43\pm 0.24)\%\,.
\label{eq:input2}
\eeq
The bound (\ref{eq:tanMH1}) also hinges on the transition from the
free quark model decay amplitude to the meson decay amplitude, as follows.
At the quark level, the dominant
contribution to $B^-\rightarrow \tau^-\,\bar{\nu}_{\tau}\,X$
comes from the exclusive quark decay
$b\rightarrow\tau^-\,\bar{\nu}_{\tau}\,c$
computed within the framework of the spectator model.
The latter works reasonably well for $B$ mesons, since the energy release
in the $b$-$c$ transition is well above $\Lambda_{\rm QCD}$ and
the typical hadronic scales ($\sim 1\,GeV$).
The next step in accuracy is to correct it
for long-distance non-perturbative effects.
In the presence of a heavy
quark, such as the bottom quark, the
leading non-perturbative corrections can be tailored with a
QCD-based operator product expansion in powers of $1/m_b$
within the context of the
heavy quark effective theory (HQET)
\footnote{See e.g. Ref.\cite{Neubert} for a
detailed review of the HQET methods.}.
This method has been worked out in detail
in Ref.\cite{Falk} to account for the semi-tauonic
$B$- meson decays, and we refer the
reader to these references for more information.
Furthermore, hard gluon exchange can be as important
as the HQET corrections,
so that in general one has to include the
${\cal O}(\alpha_s)$ short-distance
perturbative QCD effects, where $\alpha_s(m_b)\simeq 0.22$.
These effects have
been evaluated in Refs.\cite{Falk,Czarnecki2} for
the standard ($W$-mediated) amplitude.

In $2$HDM extensions of the SM, the previous analysis
must be generalized to include the HQET-type and
${\cal O}(\alpha_s)$ QCD corrections
from the $H^-$-mediated amplitude (in interference with the $W^-$ amplitude).
These contributions
have been computed in Refs.\cite{Grossman,QCDHaber},
and the bound (\ref{eq:tanMH1}) was obtained. Notice that
since the ${\cal O}(\alpha_s)$ corrections to the $W$-mediated
amplitude cancel to a large extent in $R$, one would naturally
expect that the relevant QCD corrections as far as
the $\tan\beta-M_H$ bound
is concerned should be those affecting the $H$-mediated
amplitude for the semi-tauonic $B$-decay. Notwithstanding, in practice
even this radiative effect is not too dramatic\,\cite{QCDHaber}, at least
for the ordinary QCD corrections.

The bound (\ref{eq:tanMH1}) is usually considered as very strong, for
there are no additional tree-level contributions
to $B^-\rightarrow \tau^-\,\bar{\nu}_{\tau}\,X$ aside from
$W^-$ and $H^-$ exchange. In particular,
there are no tree-level exchange of SUSY particles
(sparticles) in the MSSM. For this reason, the
upper limit (\ref{eq:tanMH1}) is usually believed to be
essentially model-independent; and at face value one
would immediately translate it to the MSSM Higgs sector
by arguing that the one-loop SUSY effects are at least as tiny as the
ordinary QCD corrections. Remarkably enough,
however, this turns out {\it not} to be the case in general,
as we shall show by explicitly computing the supersymmetric
short-distance QCD corrections (SUSY-QCD), which are expected to be
here the leading SUSY effects also\,\cite{CGJJS}. As a result, we will find
that quantum effects in the MSSM should most likely amount to
a more restrictive bound. In some cases, though, the bound
can be more relaxed, and even evanesce.

To the best of our knowledge, the potential impact of
SUSY quantum effects on semi-tauonic $B$-meson decays
has not been assessed in the literature.
However, we expect (see below) that one-loop gluino exchange can be
very important; and, indeed, we find that the
bound (\ref{eq:tanMH1}) is not as model-independent as originally thought.
It may become significantly renormalized in the MSSM, where
it has to be rephrased in a more complicated
way as a function of the SUSY parameters
\beq
(\tan\beta,M_H,\mu,\mg,\mq)\,,
\label{eq:set}
\eeq
where $\mu$ is the higgsino mixing parameter, $\mg$ is the gluino mass
and $\mq$ are the scharm and sbottom masses ($\tilde{q}=\tilde{c},\tilde{b}$).

To evaluate the quantum corrections,
we shall adopt the on-shell renormalization scheme\,\cite{BSH}.
Apart from the standard interactions
mediated by the weak gauge bosons, the Yukawa type
Lagrangian describing the charged Higgs interactions between
$b$ and $c$ quarks in the MSSM reads as follows:
\beq
{\cal L}_{Hcb}={g\,V_{cb}\over\sqrt{2}M_W}\,H^+\,\bar{c}\,
[m_b\tan\beta\,P_R+m_c\cot\beta\,P_L]\,b +{\rm h.c.}\,,
\label{eq:LcbH}
\eeq
where $P_{L,R}=1/2(1\mp\gamma_5)$ are the chiral projector operators
and $V_{cb}\simeq 0.04$ is the corresponding CKM matrix element.
This matrix element cancels out in our analysis since we
shall be concerned with the ratio (\ref{eq:ratio}).

The relevant supersymmetric parameters (\ref{eq:set}) for our
analysis are contained in the SUSY-QCD Lagrangian\,\cite{MSSM} and in the
scharm and sbottom mass matrices:
\begin{equation}
{\cal M}_{\tilde{q}}^2 =\left(\begin{array}{cc}
 {\cal M}_{11}^2 & {\cal M}_{12}^2
\\ {\cal M}_{12}^2 &{\cal M}_{22}^2 \,.
\end{array} \right)\,,
\label{eq:sqmatrix}
\end{equation}
\begin{eqnarray}
{\cal M}_{11}^2 &=&M_{\tilde{q}_L}^2+m_q^2
+\cos{2\beta}(T^3_q-Q_q\,\sin^2\theta_W)\,M_Z^2\,,\nonumber\\
{\cal M}_{22}^2 &=&M_{\tilde{q}_R}^2+m_q^2
+Q_q\,\cos{2\beta}\,\sin^2\theta_W\,M_Z^2\,,\nonumber\\
{\cal M}_{12}^2 &=&m_q\, M_{LR}^q\nonumber\\
M_{LR}^{\{c,b\}}&=& A_{\{c,b\}}-\mu\{\cot\beta,\tan\beta\}\,.
\end{eqnarray}
Since $\tilde{c}$ and $\tilde{b}$ squarks
belong to different weak-isospin multiplets, there is no $SU(2)$
correlation between the soft SUSY-breaking parameters in the two mass
matrices.

Diagonalization of ${\cal M}_{\tilde{q}}^2$ is performed
by independent rotation $2\times 2$ matrices, $R^{(\tilde{q})}$.
We will denote by $m_{\tilde{c}_1}$ ($m_{\tilde{b}_1}$) the
lightest scharm (sbottom) mass-eigenvalues.
For the sake of simplicity, we treat the two $R^{(\tilde{q})}$
assuming that the mixing angles are $\pi/4$. This is no loss of
generality, since the feature of ${\cal M}_{\tilde{q}}^2$
that really matters for our calculation
is that the off-diagonal element of the sbottom mass
matrix is non-vanishing,
so that at high $\tan\beta$ it behaves like
$m_b\,M_{\rm LR}^b\simeq -\mu\,\,m_b\tan\beta$.
The scharm mixing matrix, instead, is basically
diagonal, for $m_c/m_{\tilde{c}_1}\ll 1$  and $M_{\rm LR}^c$ is
not $\tan\beta$-enhanced.

The various contributions to the decay rate
$\Gamma (B^-\rightarrow \tau^-\,\bar{\nu}_{\tau}\,X)$
are expressed as follows:
\beq
\Gamma_{\rm B}=\Gamma_{\rm HQET}+\delta\Gamma_{\rm W,H}+\delta\Gamma_{\rm I}\,.
\label{eq:GammaB}
\eeq
Here $\Gamma_{\rm HQET}$ is the contribution from the HQET-corrected
amplitudes mediated by $W^-$, $H^-$ and interference terms
at the tree-level, and $\delta\Gamma_{\rm W,H,I}$
are the short-distance QCD and SUSY-QCD corrections.
For the semileptonic $B$-decay rate,
$\Gamma (B^-\rightarrow l^-\,\bar{\nu}_{l}\,X)$,
we have a similar formula (\ref{eq:GammaB}) but we neglect all effects
related to Higgs and interference terms.

The HQET corrections depend on a set of
parameters $(\bar{\Lambda},\lambda_1,\lambda_2)$ that connect
the $B$ and $D$ meson masses to the bottom and charm quark
masses\,\cite{Neubert,Falk}:
\beqn
m_B &=& m_b+ \bar{\Lambda}-{\lambda_1+3\,\lambda_2\over 2\,m_b}
+...\,,\nonumber\\
m_D &=& m_c+ \bar{\Lambda}-{\lambda_1+3\,\lambda_2\over 2\,m_c}+...\,.
\eeqn
This correlation between
the pole masses $m_c$ and $m_b$ is one of the main
improvements with respect to the spectator model.
The explicit form for $\Gamma_{\rm HQET}$ as a function of these parameters
is provided in Refs.\cite{Falk,Grossman}. Fortunately,
the standard QCD and SUSY-QCD contributions to $\delta\Gamma_{\rm W}$ can also
be extracted from the literature\,\cite{Falk,Czarnecki2,DHJJS}
and cancel to a large extent\footnote{The SUSY-QCD corrections to the
$W$-mediated amplitude can be derived from the work of
Ref.\cite{DHJJS}. They not only partially cancel out in the ratio $R$,
but are rather small by themselves, namely of ${\cal O}(1)\%$.
In contrast, the ordinary QCD corrections\,\cite{Falk,Czarnecki2} are of
${\cal O}(10)\%$ but cancel in $R$ to within ${\cal O}(1)\%$. } in
the ratio (\ref{eq:ratio}).

Of special relevance are the QCD and SUSY-QCD contributions
to the Higgs and interference terms, $\delta\Gamma_{\rm H,I}$. They can
be computed using the framework of Refs.\cite{QCDHaber,CD,Ricard}.
After a straightforward calculation, we arrive at
the following formulae:
\beqn
\delta\Gamma_{\rm H}&=&K\,
\int_{\rho_{\tau}}^{(1-\sqrt{\rho_c})^2}\,dt\ \ H(t)\,
\left(1-{\rho_{\tau}\over t}\right)^2\,
{t\,\rho_{\tau}\,\tan^2\beta\over\xi^2}
\,(\delta a+\delta b)\,,\nonumber\\
\delta\Gamma_{\rm I}&=&-K\,
\int_{\rho_{\tau}}^{(1-\sqrt{\rho_c})^2}\,dt\ \ I(t)\,
\left(1-{\rho_{\tau}\over t}\right)^2\,
{\rho_{\tau}\over\xi}
\,\left[\right.\delta a+\delta b
 +\,\sqrt{\rho_{c}}\,\left(\delta a-\delta b\right)\left.\right]\,,
\label{eq:GammaHI}
\eeqn
with
\beq
K={G_F\,m_b^2\,\tan^2\beta\over 4\,\sqrt{2}\,\pi^2}\,,\ \ \
\rho_c=m_c^2/m_b^2\,,\ \ \ \rho_{\tau}=m_{\tau}^2/m_b^2\,,
\ \ \ \xi=M_{H}^2/m_b^2\,.
\label{eq:coeffs}
\eeq
We have introduced
\beq
H(t)=\,\Gamma_{\rm bcS}(\rho_c,t;2,0,1)\,,\ \ \ \
I(t)=\Gamma_{\rm bcS}(\rho_c,t; 2,-2\,\sqrt{\rho_c},1)\,,
\eeq
where $\Gamma_{\rm bcS}(\rho_c,t;c_1,c_2,c_3)$ is an appropriate
(tree-level) function
defined in Ref.\cite{QCDHaber};
it is related to the decay rate $b\rightarrow c S$
into a virtual scalar $S=H^-,G^-$.
Here $G^-$ is a Goldstone boson contribution,
for the calculation is carried out in
the convenient setting of the Landau gauge.
Furthermore,
$\delta a, \delta b$ in eq.(\ref{eq:GammaHI})
contain the ordinary QCD plus SUSY-QCD corrections to
the effective couplings $a=a_H+a_G$ and $b=b_H+b_G$
standing in the interaction Lagrangian of $S$ with charm and bottom quarks:
\beq
{i\,g\,m_b\,\tan\beta\,V_{cb}\over
2\,\sqrt{2}\,M_W}\,\bar{c}(a+b\,\gamma_5)\,b\,S\,.
\label{eq:abS}
\eeq
The standard QCD corrections
$\delta a^{\rm QCD}$ and $\delta b^{\rm QCD}$
can be obtained by adapting the
results of Ref.\cite{CD} whereas the SUSY-QCD corrections
follow after straightforward modification of
the form factors $G_L, G_R$ of Ref.\cite{Ricard}:
\beqn
\delta a^{\rm SUSY-QCD}=G_R+{\sqrt{\rho_c}\over\tan^2\beta}\,G_L\,,\nonumber\\
\delta b^{\rm SUSY-QCD}=G_R-{\sqrt{\rho_c}\over\tan^2\beta}\,G_L\,.
\label{eq:GLGR}
\eeqn
In the limit
of large $\tan\beta$,
\beq
\delta a^{\rm SUSY-QCD}\simeq\ \delta b^{\rm SUSY-QCD}
\simeq\ G_R = H_R+{\delta m_b\over
m_b} +\frac{1}{2}\,\delta Z_L^c+\frac{1}{2}\,\delta Z_R^b\,,
\label{eq:GR}
\eeq
where $H_R$ is a vertex form factor and the remaining terms are suitable
mass and wave-function renormalization
counterterms in the on-shell scheme, and can be readily identified
from the work of Refs.\cite{Ricard,GJS}.

Collecting all the  pieces from the RHS of eq.(\ref{eq:GammaB}),
we may now perform the numerical analysis of the
ratio $R$ -- Cf. Figs.1-3 and Table 1.
We fix the error bars for the
HQET parameters as in Ref.\cite{QCDHaber}; and
to account for the uncertainties associated to ${\cal O}(\alpha_s^2)$
corrections, we have also varied the renormalization scale such that
$0.20\leq \alpha_s\leq 0.36$.
As a first step in our numerical analysis (Cf. Figs.1a and 3), we have
carefully checked that we are able to recover the non-supersymmetric
results\,\cite{QCDHaber,Falk,Grossman}. Indeed,
upon disconnecting the SUSY terms, we have verified
(with the help of MINUIT) that we accurately
reproduce the numerical results obtained for $R$ as
a function of $r\equiv \tan\beta/M_H$ (Cf. Fig.1 of Ref.\cite{QCDHaber});
in particular, we recover the bound (\ref{eq:tanMH1}) based on the
inputs (\ref{eq:input1})-(\ref{eq:input2}).

At present, the experimental situation has changed a little.
For example, a recent
ALEPH measurement yields\,\cite{Anna}
\beq
BR(B^-\rightarrow \tau^-\,\bar{\nu}_{\tau}\,X)=(2.72\pm 0.34)\%\,,
\label{eq:input3}
\eeq
which is slightly more tight\footnote{We point out the two experimental
values since the older one, eq.(\ref{eq:input1}), is an average of
previous ALEPH and L$3$ measurements\,\cite{ALEPHL3}, whilst the
new one, eq.(\ref{eq:input3}), is an average
of measurements of only the ALEPH Collaboration based on two
different experimental techniques\,\cite{Anna}.}. However, also the
inclusive semileptonic branching ratio has changed slightly.
On the one hand, the LEP electroweak working group uses\,\cite{LEPEWW}
\beq
BR(B^-\rightarrow l^-\,\bar{\nu}_{l}\,X)=(11.2\pm 0.4)\%\,,
\label{eq:input4}
\eeq
and on the other hand the CLEO/ARGUS and L$3$ results suggest
a lower value\,\cite{LEPEWW}
which brings the average closer to (\ref{eq:input2}).
In view of this situation, and since we wish to make clear that our SUSY
effects are potentially ``real'', i.e. that they are not just an artifact
associated to the change of the experimental inputs,
we shall first of all normalize our analysis
with respect to the same inputs (\ref{eq:input1})-(\ref{eq:input2}) used in
Ref.\cite{QCDHaber} and thus present our results (Figs.1-3) in this framework.
Notice that, in the SUSY case, the analysis cannot be
strictly formulated in
terms of the single parameter $r$, but rather as a function of $\tan\beta$,
$M_H$ and the rest of
parameters (\ref{eq:set}). Still, limits on an
effective $r$ can be given
after fully exploring the parameter space. These are given
in Table 1, where we exhibit in a nutshell our final
numerical results on the $\tan\beta-M_H$ bound. We reserve
the last column of that table for the results obtained by using
the most recent LEP data, i.e.
eqs.(\ref{eq:input3})-(\ref{eq:input4}).
By this procedure we have fitted the relevant part of the
physical boundary of the
$(\tan\beta,M_H)$-space to the linear
form $\tan\beta=r_{\rm max}\,M_H$,  where $r_{\rm max}$ is the
maximum ``effective slope'' compatible with the sparticle mass parameters
given in Table 1. In this way we can easily compare our SUSY results with
the general (Type II) $2$HDM bound
(\ref{eq:tanMH1}). It should be emphasized that a good
local linear regression in the SUSY case is possible (Cf. Fig.3) because
the ratio $R$ has low sensitivity to the squark and gluino masses
in the few hundred $GeV$ range, as it
is borne out in Figs.2a-2b -- see further comments below.

In the following we analyze things in more detail.
For fixed  $M_H=120\,GeV$, the plot of $R$, eq.(\ref{eq:ratio}),
as a function of $\tan\beta$ is shown in Fig.1a for $\mu<0$
and in Fig.1b for $\mu>0$. The shaded region
gives the experimental band at $1\,\sigma$ as determined
from eqs.(\ref{eq:input1})-(\ref{eq:input2}).
The SUSY-QCD effects in Figs.1a-1b are computed for the (approximate)
present bounds on sparticle masses, namely $\mg=150\,GeV$ and
$m_{\tilde{b}_1}=m_{\tilde{c}_1}=150\,GeV$ (Scenario A in Table 1).
We have also fixed $A_c=A_b=300\,GeV\equiv A$, but
the dependence on this parameter is not important at high $\tan\beta$.
A most interesting parameter is $\mu$. It is plain from Figs.1a-1b
that both the sign and size of $\mu$ are material;
indeed, the larger
is $|\mu|$ (for $\mu<0$) the steepest is the ascent of $R$
into the experimental band
and so the narrower is the preferred interval of high $\tan\beta$ values.
On the contrary, for smaller and smaller $|\mu|$ the SUSY effect dies away.
In all figures where a definite $\mu<0$ is to be chosen,
the value $\mu=-80\,GeV$ (Scenario A) is taken, except in Fig.1b where
the case $\mu>0$ is addressed in detail.
It is easy to see from the
structure of the chargino mass matrix in the
higgsino-gaugino variables\,\cite{MSSM}
that, in the high $\tan\beta>10$ region,
$\mu=-80\,GeV$ is the minimum allowed value
of $|\mu|$  compatible with the
LEP $1.61$ phenomenological bound
$m_{\chi^{\pm}_1}\stackM 80\,GeV$\,\cite{RMiquel}.
The latter is the strongest phenomenological mass limit on charginos
available from LEP, and corresponds to the so-called
neutralino LSP scenario.

Due to the variation of the HQET parameters and $\alpha_s$ in the
aforementioned ranges,
our results are not single curves but `` beam curves''.
For a better understanding, in Fig.1a we have simultaneously plotted,
as a function of $\tan\beta$, the beam curves for:
\begin{itemize}
\item{(i)} The fully SUSY-QCD
corrected ratio $R$, eq.(\ref{eq:ratio}), which we call
$R_{SUSY}$, including all effects present in eq.(\ref{eq:GammaB});
\item{(ii)} The Higgs-corrected ratio $R$, denoted $R_H$,
with HQET and ordinary QCD corrections but without SUSY-QCD effects;
\item{(ii)} The ratio $R$ without Higgs effects, i.e. the
(so-called) standard model (SM) prediction, computed with only the
$W$-mediated amplitude including HQET and ordinary QCD corrections.
It is represented in Fig.1a by the narrow dotted band defined by
$R_{SM}=0.22\pm 0.02$.
\end{itemize}
{}From Fig.1a it is patent that the SM prediction lies
tangentially below the experimental range, specifically
$1\,\sigma$  below the central value of the
experimental band. Admittedly,
this ``$R$ anomaly'' is not that
serious and varies a bit depending on the data set used.
In any case a useful bound on multiple Higgs extensions of the SM can be
derived.
It is evident from Fig.1a that charged Higgs effects
go in the right direction; for they shift the theoretical result entirely
into the experimental range, to the extent that
$R_H$ may even overshoot the upper experimental
limit at sufficiently high $\tan\beta$. To prevent this
from happening, the bound
(\ref{eq:tanMH1}) must be imposed\,\cite{QCDHaber}.
Similarly, if the charged Higgs is a SUSY Higgs, there are
additional SUSY effects that may substantially alter the picture both
quantitatively and qualitatively.
Indeed, a most vivid SUSY impinge on $R$ occurs for
$\mu<0$ which
triggers a sudden ``re-entering'' of the theoretical ratio $R_{SUSY}$
into the experimental band at an earlier value of
$\tan\beta$ and for a sharper range than in the $R_H$ case.

We point out that
the sign $\mu>0$, although it is not the most suited one
for $R$ (Cf. Fig.1b), it cannot be convincingly excluded,
and it may even hide some surprises.  To start with, we observe that
the $\mu>0$ beam curves overlap with the SM band all the way up
to $\tan\beta\stackM 40$. Nevertheless, for $\tan\beta>40$ the beams
behave very differently, depending on the value of $\mu$;
to wit: i) If $\mu>80-90\,GeV$, they
quickly run away the experimental band from below;
ii) If $\mu<20\,GeV$, they
bend back into the experimental range past the $R_H$ limit;
iii) Finally, if $20\,GeV\stackm\mu\stackm 80\,GeV$, they
spread very widely, mainly because of
the variation with $\alpha_s$.
In the first case, $\mu>0$ becomes excluded at very high $\tan\beta$;
in the second case,
the bound (\ref{eq:tanMH1}) is violated
since larger values of $\tan\beta$ are allowed for a given $M_H$;
and in the third case, remarkably enough, the beam curves
(partly) overlap  all the time with the experimental
region until the perturbative limit $\tan\beta\stackm 60$
is already met. Therefore,
for $20\,GeV\stackm\mu\stackm 80\,\,GeV$ a dramatic qualitative
change occurs: the bound is fully destroyed, i.e.
at the $2\,\sigma$ level there is no bound at all!.
However, small values of $|\mu|$ are not recommended
by present LEP data, as advertised before,
and in this sense $\mu>0$ values in the previous interval
might already be excluded by LEP.

{}From the point of view of the ``$R$ anomaly'',
the sign $\mu<0$ becomes strongly preferred since, then, there
always exists a high $\tan\beta$ interval where
all the beam curves rush into the experimental band for any
value of $|\mu|$.
In this case, compatibility with $b\rightarrow s\,\gamma$ requires
$A_t>0$\,\cite{Ng}. Hence at present the combined
status of neutral and charged $B$-meson decays points to
the signs $\mu<0$ and $A_t>0$. This feature does
not depend on the values of the other SUSY parameters (\ref{eq:set}),
as it is confirmed in Figs. 2a-2b
where we explore the dependencies on the sbottom and gluino masses
for fixed $\tan\beta$. The evolution
with $m_{\tilde{b}_1}$ shows a slow
decoupling (Fig.2a) while the dependence on $m_{\tilde{g}}$ is such that,
locally, the SUSY-QCD corrections
slightly increase  with $m_{\tilde{g}}$ (Cf. Fig.2b)
and eventually decouple (not shown).
However, the decoupling rate turns out to be
so slow that one may reach $m_{\tilde{g}}\sim 1\,TeV$
without yet undergoing dramatic suppression.
Finally, the evolution with the scharm masses is very mild and it
is not displayed.

It is instructive to isolate the leading source
of SUSY-QCD effects.
It originates from a
(finite) bottom mass renormalization effect in the form
factor $G_R$\,\cite{Ricard,CGJJS}.
Specifically, this effect is contained in eq.(\ref{eq:GR}) as follows:
\beqn
\left({\delta m_b\over m_b}\right)_{\rm SUSY-QCD} &=&
C_F\,{\alpha_s\,\over 2\pi}\,m_{\tilde{g}}\,M_{LR}^b\,
I(m_{\tilde{b}_1},m_{\tilde{b}_2},m_{\tilde{g}})\,+...\nonumber\\
&\simeq& -{2\alpha_s\over 3\pi}\,m_{\tilde{g}}\,\mu\tan\beta\,
I(m_{\tilde{b}_1},m_{\tilde{b}_2},m_{\tilde{g}})\,+... \,,
\label{eq:dmbQCD}
\eeqn
where $C_F=(N_c^2-1)/2\,N_c=4/3$ is a colour factor.
The last result holds for sufficiently large $\tan\beta$. We have introduced
the positive-definite function
\beq
I(m_1,m_2,m_3)=
{m_1^2\,m_2^2\ln{m_1^2\over m_2^2}
+m_2^2\,m_3^2\ln{m_2^2\over m_3^2}+m_1^2\,m_3^2\ln{m_3^2\over m_1^2}\over
 (m_1^2-m_2^2)\,(m_2^2-m_3^2)\,(m_1^2-m_3^2)}\,.
\label{eq:I123}
\eeq
Formally, eq.(\ref{eq:dmbQCD}) describes the same one-loop
threshold effect from
massive particles that one has to introduce to correct
the ordinary massless contributions
(i.e. to correct the standard QCD running bottom quark mass) in
SUSY GUT models\,\cite{SO10}.
As an aside, we point out that the so-called light gluino scenario is not
favoured
in our case, since eq.(\ref{eq:dmbQCD}) vanishes for $m_{\tilde{g}}=0$.
However, there is another potentially large $\tan\beta$
effect\,\cite{SO10}, the one due
to chargino-stop diagrams\,\cite{Referee}.
This is typically less than the gluino
diagram, and it is reasonable to neglect it in most of parameter space.
Although it is true that for vanishing gluino mass it could dominate the
large $\tan\beta$ effects, notice that
the light gluino scenario is nowadays essentially dead. Recent LEP analyses
do exclude light gluinos below $6.3\,GeV$\,\cite{LEPgluino}.
For gluino masses as in  Table 1, compatibility with
$b\rightarrow s\,\gamma$ (see below) renders that effect
subleading in all cases.

In Fig.3 we display the results of our analysis in
the $(\tan\beta,M_H)$-plane for Scenario A and
inputs (\ref{eq:input1})-(\ref{eq:input2}).
As stated, we concentrate
on the case $\mu<0$. Recall that, in the MSSM, the
lowest allowed charged Higgs mass is $M_H\stackM 100\,GeV$
since it is correlated with the present LEP bound on the
CP-odd Higgs mass, $M_{A^0}\stackM 60\,GeV$.
At the $2\,\sigma$ level, the
allowed region by the SUSY-corrected ratio $R_{SUSY}$ is the
big shaded area on the left upper part of Fig.3.
In contrast, at the $1\,\sigma$ level
the permitted area is much smaller,
and it is represented by that slice of the big
shaded area limited by the two thin solid lines.
Of course, lower segments of $\tan\beta$
are also allowed at $1\,\sigma$ but
they do not entail any
improvement at all with respect to the SM.
Hence if we just concentrate on the high $\tan\beta>30$ region,
it turns out that at $1\,\sigma$
there exists only a narrow range of optimal $\tan\beta$ values
for any given $M_{H}$. This was already evident from Fig.1a where
$M_H=120\,GeV$.
If we would now superimpose the perturbative
limit ($\tan\beta\stackm 60$) we would find that the highest allowed
charged Higgs mass in Fig.3 is rather small:
$M_{H}<190\,GeV$.
At $2\,\sigma$, however, we have seen that
the situation is far more permissive and
one cannot place that bound; yet the allowed  area  by $R_{SUSY}$
at $2\,\sigma$ is significantly smaller than the one allowed (at the same
confidence level) by the non-supersymmetric ratio $R_H$ (see dashed line
in Fig.3).
A good linear approximation to the SUSY-corrected $2\,\sigma$ boundary
is possible in the $(\tan\beta,M_H)$ window of Fig.3.
It corresponds to an ``effective slope''
of $r_{\rm max}=0.44\,GeV^{-1}$.

{}From a recent analysis\,\cite{ConwayRoy} of
$\tau$-lepton physics at the Tevatron,
based on the {\it non}-observation of the decay
$t\rightarrow H^+\,b$ followed by $H^+\rightarrow \tau^+\,\nu_{\tau}$,
it is possible to find a different (high-energy) exclusion plot in the
$(\tan\beta,M_H)$-plane.
The latter is represented, at the $2\,\sigma$ level, by the cross-hatched area
in the low right corner of Fig.3\,\footnote{We remark
that the $(\tan\beta,M_H)$ exclusion plot from
 $t\rightarrow H^+\,b$  is not fully watertight
since it should be revised in the light of
potentially important supersymmetric quantum effects recently
computed in the context of the MSSM\,\cite{CGJJS}.}.
Interestingly enough, however,
it turns out that the excluded region of
the $(\tan\beta,M_H)$-plane obtained from SUSY-corrected
$B$-meson decays is the most stringent one and
basically overrides the other
exclusion plots,
as can be appraised in Fig.3.

A final remark is in order. We have stated at the beginning that
the decay $b\rightarrow s\,\gamma$ plays an important role in constraining
the MSSM parameter space. Therefore, it is necessary to check explicitly
the compatibility between the $b\rightarrow s\,\gamma$ constraints and the
ones from semi-tauonic $B$-decays\,\cite{Referee}.
This can most easily be performed using the
formula (see the extensive literature\,\cite{Ng} for details)
\beq
BR (b\rightarrow s\,\gamma)\simeq BR (b\rightarrow c\,e\,\bar{\nu})\,
{(6\,\alpha_{\rm em}/\pi)\,\left(\eta^{16/23}\,A_{\gamma}+C\right)^2\over
I(m_c/m_b)\,\left[1-\frac{2}{3\pi}\,\alpha_s(m_b)f_{\rm QCD}(m_c/m_b)\right]}
\label{eq:bsg}
\eeq
where $A_{\gamma}=A_{\rm SM}+A_{H^-}+A_{\chi^-}$ stands for the sum of the
SM, charged Higgs and chargino-stop amplitudes, respectively. The contribution
from a SUSY-QCD amplitude is in this case generally smaller.
(Notice that when using eq.(\ref{eq:bsg}) one should make allowance for
possible additional corrections of order $30\%$ stemming
from higher order QCD effects not included in it\,\cite{Ng}.)
The $b\rightarrow s\,\gamma$ consistency check is
necessary because one may worry whether
at large $\tan\beta$ (the regime that we have favoured in our study
of semi-tauonic $B$-decays)
and for sizeable $A_t$ the chargino
amplitude might be enhanced and perhaps overshoot
the CLEO bound\,\cite{CLEO} in the other direction, i.e.
it could grow to the point of overcompensating
the SM plus charged Higgs contribution. However, we have explicitly checked
in all cases
that upon using the same input parameters as in the present analysis, the CLEO
bound can be respected.  For definiteness in our presentation, we
consider the following set of inputs:
$M_{H^\pm}=120\,GeV$, $\mu=-80\,GeV$ and $\tan\beta=30-40$.
We then find two possible types of solutions: namely,  either the two
stops are relatively heavy (roughly degenerate at about $300\,GeV$)
and $A_t$ remains bounded within any of the  approximate intervals
$(10,60)\,GeV$ and $(100,150)\,GeV$; or
another possibility is that one of the stops
is relatively light (for example just above
the present approximate LEP bound: $m_{\tilde{t}_1}>65\,GeV$)
and the other one is very heavy (we take it $m_{\tilde{t}_2}=1\,TeV$).
In this case $A_t$ is forced to lie
in the approximate intervals $(-400,-200)\,GeV$ and $(-100,+20)\,GeV$.
For any of the two possible type of solutions,
the SUSY electroweak contributions to $R$  can be estimated
(from the work of Ref.\cite{SO10}) to be
subleading (in most cases below $10\%$) as compared to the SUSY-QCD effects.
Therefore, we conclude that in general the results presented in this paper
should not change significantly after computing the rest of the MSSM
corrections to the semi-tauonic $B$-decay.

To summarize, we have assessed the impact of the SUSY-QCD short-distance
effects on the physics of the semi-tauonic inclusive $B$-meson
decays within the framework of the MSSM. A regime of
large $\tan\beta>30$ is singled out. In this regime, the $\mu>0$ case
with $\mu>80\,GeV$ is ruled out; however,
if $\mu\stackm 80\,GeV$ is allowed, then there could be
no $\tan\beta-M_H$ bound at all, but this possibility seems to be
unfavoured by recent LEP exlusion data on chargino production.
On the other hand, for the most
likely case $\mu<0$,
the SUSY effects further restrict the allowed region in the
$(\tan\beta,M_H)$-plane as compared to eq.(\ref{eq:tanMH1}).
A clear-cut r\'esum\'e of our  $\mu<0$ results
is conveniently displayed in Table 1.
Using the present day sparticle mass limits
and the recent LEP input data
on $B$-meson decays
-- i.e. Scenario A (ii) in Table 1 -- we
have at the $1\,\sigma$\, ($2\,\sigma)$ level:
\beq
\tan\beta<0.40\ (0.43)\,\ (M_H/{\rm GeV})\,.
\label{eq:tanMH2}
\eeq
Since $M_H\geq 100\,GeV$ in the MSSM, it follows that the SUSY effects
compel the maximum allowed value of $\tan\beta$ to be at least $9$ units
smaller
than it was allowed by the previous
bound, eq.(\ref{eq:tanMH1}), i.e. in general
$r$ receives a SUSY correction over $-15\%$.
We have also considered a situation (Scenario B) where gluinos are
kept at the current phenomenological mass limit while the (lightest) sbottom
is twice as much heavier than in Scenario A.
This is the maximum conspiracy against our
bound for these mass ranges, and yet the result for $r_{\rm max}$ varies less
than
$7\%$. Finally, C and D in Table 1 reflect future scenarios
characterized by
large squark and gluino masses as well as a substantially improved limit for
the higgsino mass parameter. We wish to emphasize that,
for $|\mu|\stackM 150\,GeV$, $r_{\rm max}$
is already essentially saturated in the value of $|\mu|$,
i.e. larger values do not appreciably modify $r_{\rm max}$.
Notice that the leading effect (\ref{eq:dmbQCD}) does not decouple
when the masses of the sparticles involved in it
are scaled up by keeping their ratios fixed.
This is verified in Table 1 where we see that scenarios
A and C give essentially the same result\,\cite{Referee}.

Therefore, our results look fairly stable within the
phenomenologically interesting portion of the parameter space (\ref{eq:set})
and should be considered as rather general in the context of the MSSM.
We have also found that
at present the information
on the parameter space $(\tan\beta,M_H)$ as collected from $B$-meson decays
is more restrictive
than the one from top quark decays. Clearly, knowledge from both
low-energy and high-energy data can be very useful
to better pinpoint in the future the physical boundaries
of the MSSM parameter space.
Alternatively, if the two approaches would converge to
a given portion of that parameter
space, one could claim strong indirect evidence of SUSY.


{\bf Acknowledgements}:
\noindent
J.S. thanks Y. Grossman for early discussions on this subject.
Useful conversations with  E. Bagan, D.P. Roy and J. Soto are
gratefully acknowledged.
He is also indebted  to  A. Pascual
and M. Bosman for his interest and for providing
fresh information on the experimental status of $B$-meson decays.
Finally, we thank L. Mir for handing us the last LEP sparticle bounds.
This work has been partially supported by CICYT
under project No. AEN95-0882.


\vspace{0.1cm}
\hyphenation{ty-pi-cal}
\vspace{1cm}
\begin{center}
\begin{Large}
{\bf Figure Captions}
\end{Large}
\end{center}
\begin{itemize}
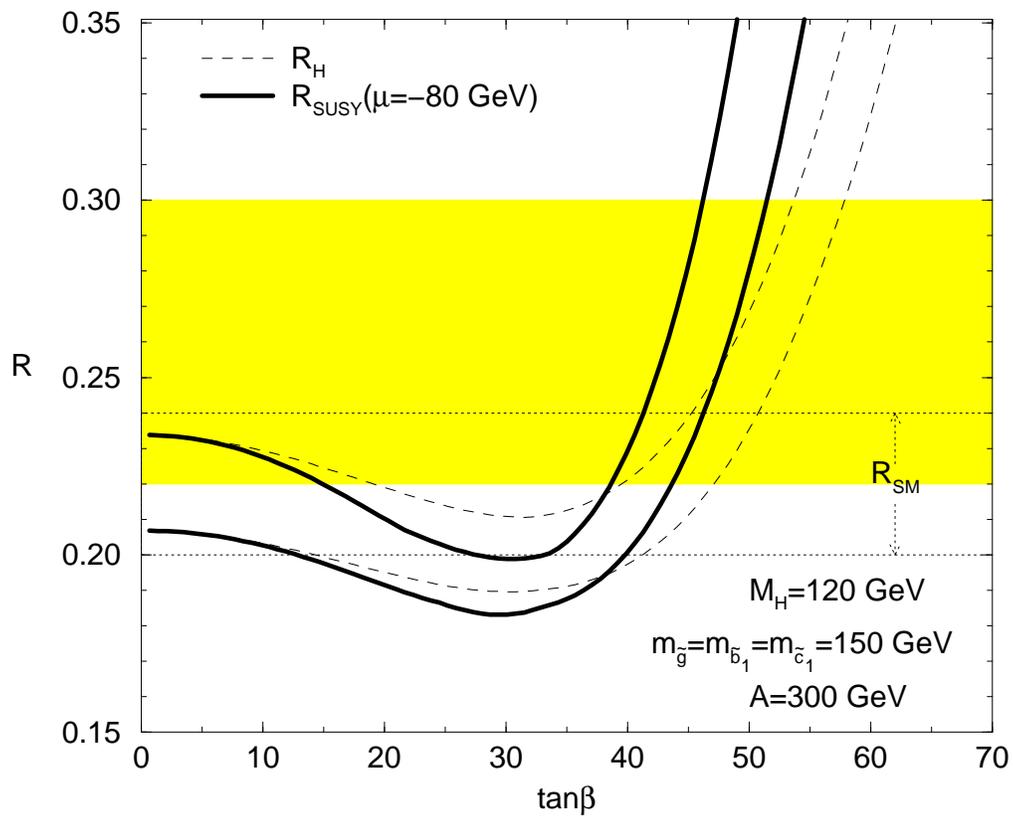
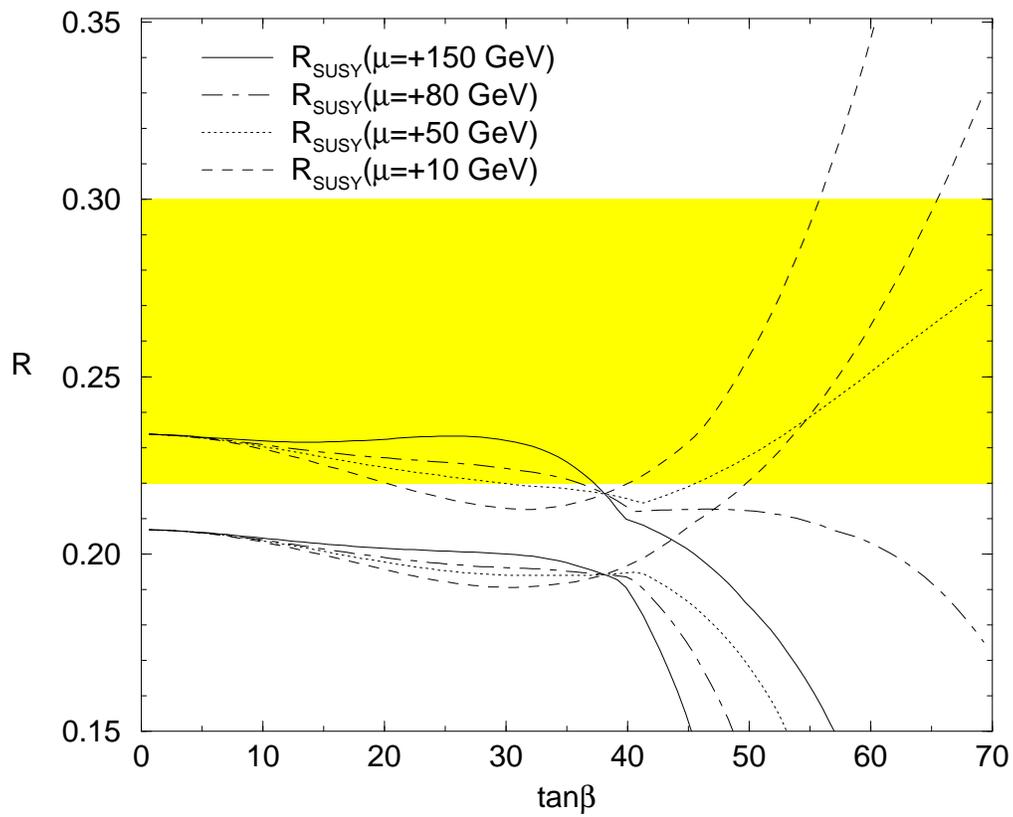
\item{\bf Fig.1}  {\bf (a)} The SUSY-QCD corrected
ratio $R_{SUSY}$, eq.(\ref{eq:ratio}), as a
function of $\tan\beta$, for $\mu=-80\,GeV$ and given values of the other
parameters (\ref{eq:set}). The HQET and $\alpha_s$ parameter ranges are as
in Ref.\cite{QCDHaber}.
Also shown are the SM result, $R_{SM}$,
(dotted band) and the Higgs-corrected
result without SUSY effects, $R_H$.
The shaded band is the experimental measurement at the $1\,\sigma$ level
as given by eqs.(\ref{eq:input1})-(\ref{eq:input2});
{\bf (b)} As in (a), but for $\mu=+10,+50,+80, +150\, GeV$.

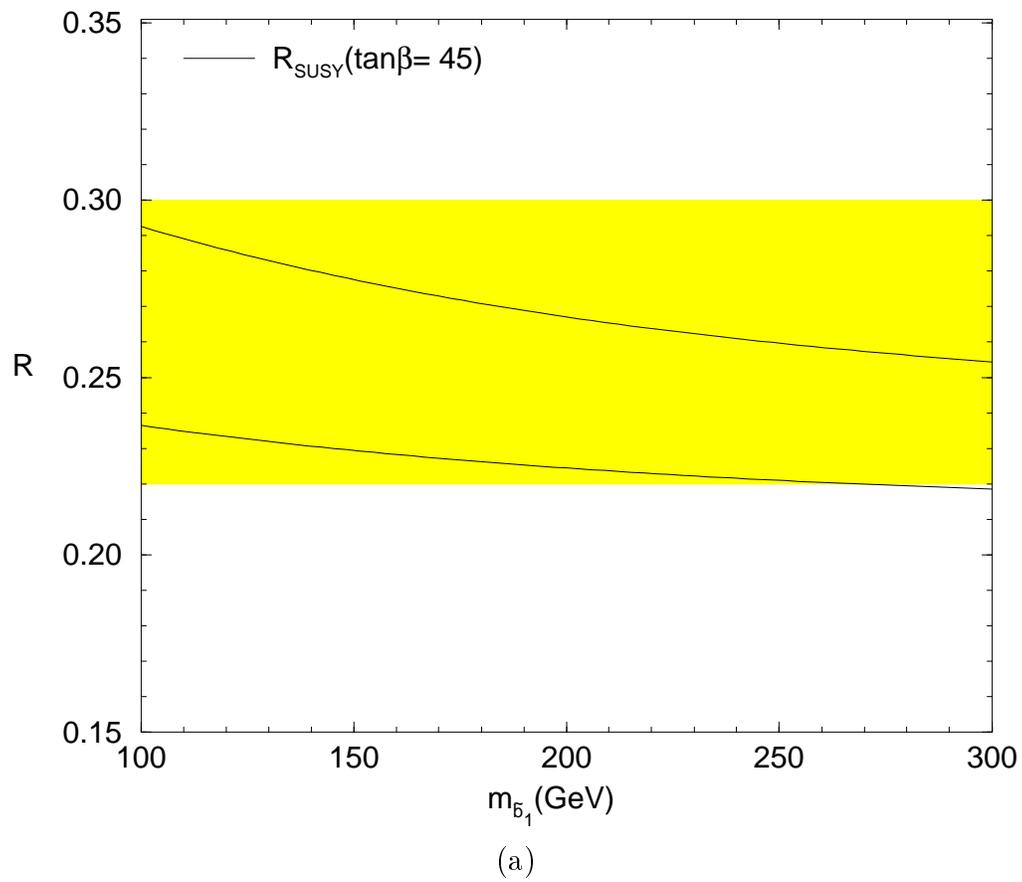
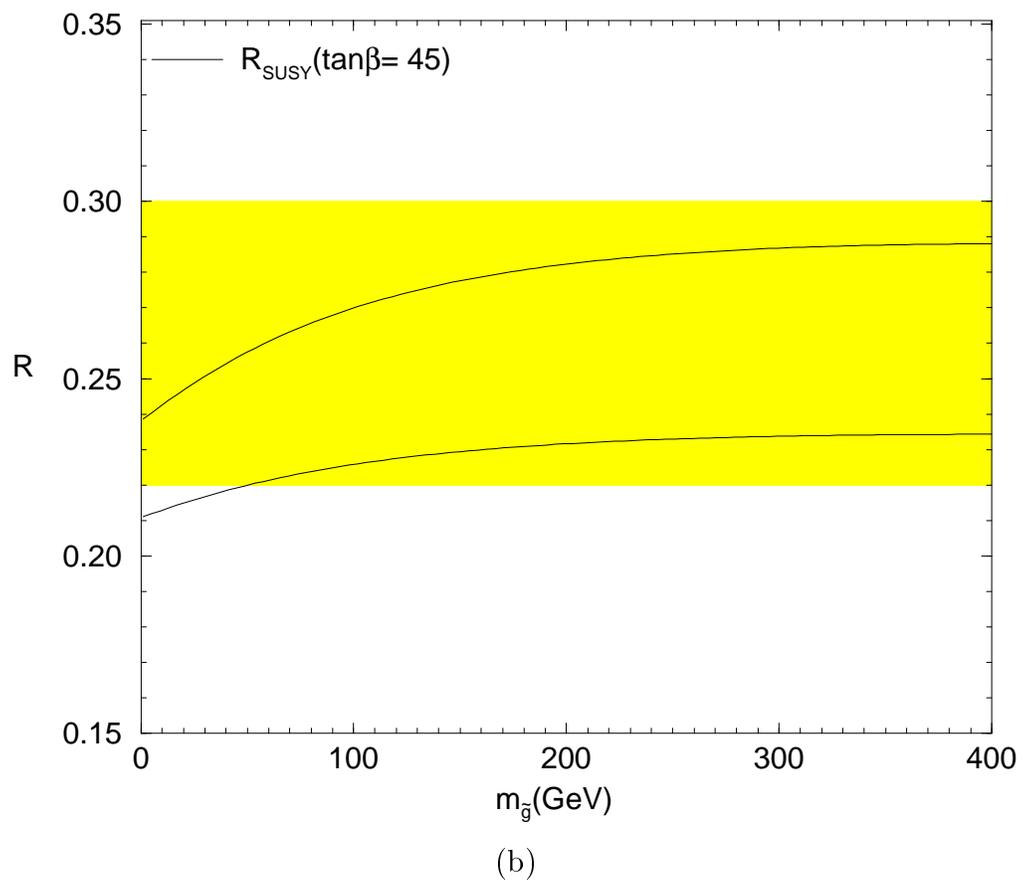
\item{\bf Fig.2} {\bf (a)} Dependence of $R_{SUSY}$
upon the lightest sbottom mass for $\tan\beta=45$.
Remaining inputs as in Fig.1a.
{\bf (b)} $R_{SUSY}$ as a function of the gluino mass,
and the rest of inputs as in (a).

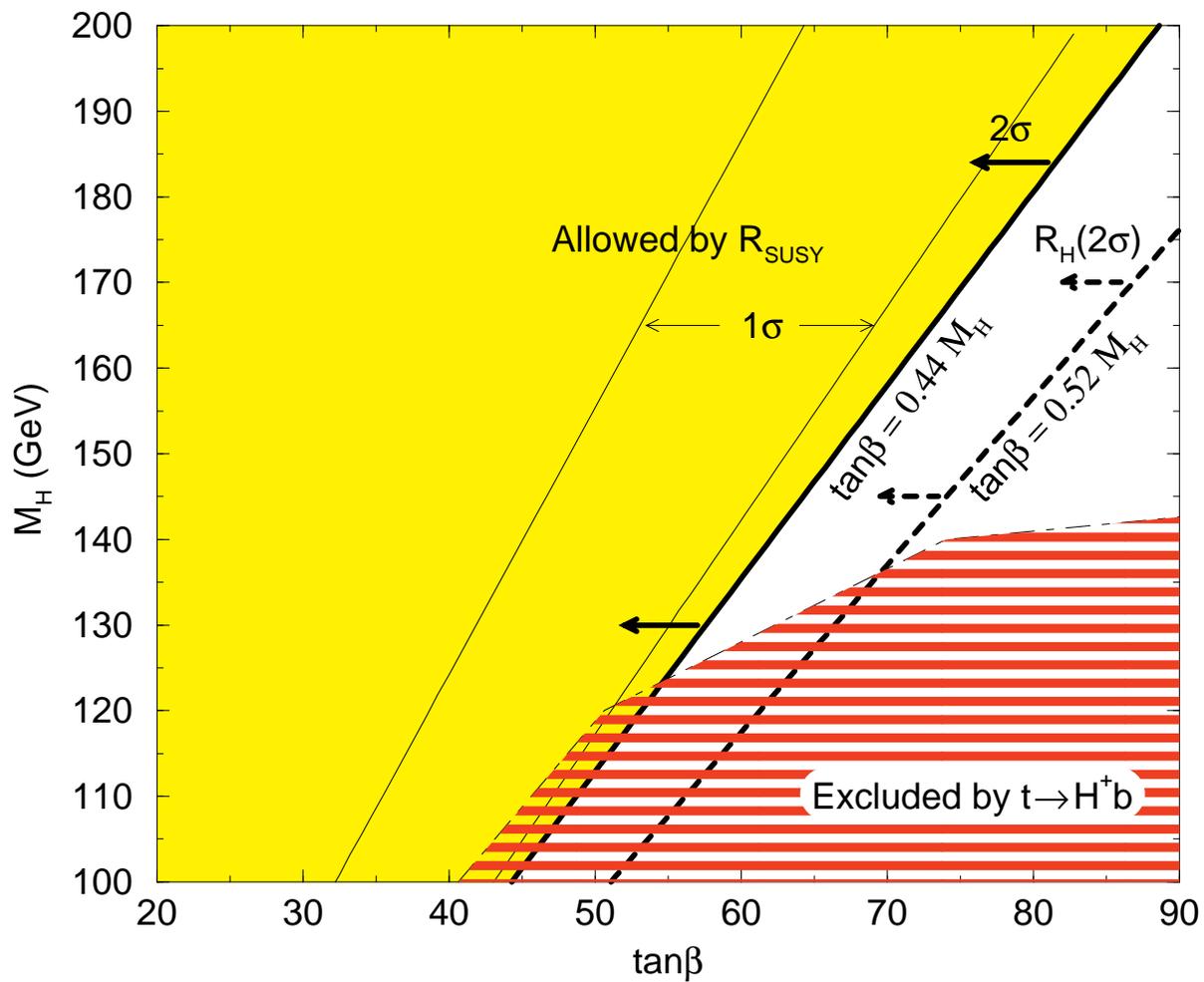
\item{\bf Fig.3} Allowed region in the $(\tan\beta,M_{H^\pm})$-plane.
Direct LEP limits on $M_{A^0}$
already imply $M_{H^\pm}\stackM 100\,GeV$ in the
MSSM. The shaded region limited by
the bold solid line is allowed at the $2\,\sigma$ level by
$R_{SUSY}$ for the same fixed parameters as in Fig.1a. The narrow subarea
between the thin solid lines is  permitted at $1\,\sigma$ only.
(The larger is $|\mu|$ the narrower is this area.) In contrast,
the allowed region at $2\,\sigma$  by the non-supersymmetric
calculation, $R_H$, is the one placed above the dashed line.
Also shown is the region excluded (at $2\,\sigma$) by the
non-observation of $t\rightarrow H^+\,b$ at the Tevatron.

\end{itemize}

\vspace{0.5cm}
\newpage
\begin{center}
\begin{Large}
{\bf Table Caption}
\end{Large}
\end{center}
\begin{itemize}
\item{\bf Table I.}\ Effective $\tan\beta/M_H<r_{\rm max}$ bound
for four $\mu<0$ scenarios: A) corresponds to
the (approximate) present day mass limits on sparticles; B)
is defined by the
combination of sparticle masses giving the worst possible bound on $r_{\rm
max}$;
finally,  C,D) reflect the situation for future sparticle mass limits.
In the four cases, we show the $1\,\sigma$ ($2\,\sigma$) upper bounds on
$r_{\rm max}$ for the two sets of inputs:
 (i) eqs.(\ref{eq:input1})-(\ref{eq:input2})
and (ii) eqs.(\ref{eq:input3})-(\ref{eq:input4}).

\end{itemize}

\begin{center}
{\bf Table 1}
\end{center}

\thispagestyle{empty}
\begin{tabular}{c||c|c|c|c|c||} \cline{2-6}
   &  $\mu\;(GeV)$  &  $m_{\tilde{g}}\;(GeV)$  &
      $m_{{\tilde{b}}_1}\;(GeV)$  &
      \multicolumn{2}{c||}{$r_{max}\;(GeV^{-1})\;\;1\sigma\;(2\sigma)$}
      \\ \cline{5-6}
   &  &  &  &  (i)  &
 (ii)  \\ \hline
 A & -80 & 150 & 150 & 0.42 (0.44) & 0.40 (0.43) \\ \hline
 B & -80 & 150 & 300 & 0.45 (0.47) & 0.43 (0.45) \\ \hline
 C & -150& 300 & 300 & 0.42 (0.44) & 0.40 (0.42) \\ \hline
 D & -300& 400 & 400 & 0.40 (0.42) & 0.39 (0.41) \\ \hline
\end{tabular}


\begin{thebibliography}{9999}
\bibitem{MSSM}
H. Nilles, {\it Phys. Rep.} {\bf 110} (1984) 1;
 H.E. Haber and G. Kane, {\it Phys. Rep.}
{\bf 117} (1985) 75.
 \bibitem{WdeBoer}
W.de Boer, A. Dabelstein, W. Hollik, W. M\"osle, U. Schwickerath,
{\it Updated global fits of the SM and the MSSM to electroweak precision data},
preprint IEK-KA.96-07 [hep-ph/9609209].
\bibitem{Barger}
V. Barger, M.S. Berger, R.J.N. Phillips,\, {\it Phys. Rev. Lett.}\,
{\bf 70} (1993) 1368; A. K. Grant,\, {\it Phys. Rev.}\,
{\bf D 51} (1995) 207.
\bibitem{Ng}
S. Bertolini, F. Borzumati, A. Masiero, G. Ridolfi, \, {\it Nucl. Phys.}\,
{\bf B 353} (1991) 591;
R. Barbieri, G.F. Giudice,\, {\it Phys. Lett.}\, {\bf B 309} (1993) 86;
R. Garisto, J.N. Ng,\, {\it Phys. Lett.}\, {\bf B 315} (1993) 372;
M. A. Diaz,\, {\it Phys. Lett.}\, {\bf B 322} (1994) 207;
F. Borzumati, \, {\it Z. Phys.}\, {\bf C 63} (1994) 291;
S. Bertolini, F. Vissani, \,{\it Z. Phys.}\, {\bf C 67} (1995) 513;
M. Carena, C.E.M. Wagner,\, {\it Nucl. Phys.}\, {\bf B 452} (1995) 45;
R. Rattazzi, U. Sarid,\,
{\it Phys. Rev.} {\bf D 53} (1996) 1553.
\bibitem{CGJJS}
J.A. Coarasa, D. Garcia, J. Guasch, R.A. Jim\'enez, J. Sol\`a,
{\it Quantum effects on $t\rightarrow H^+\,b$ in the MSSM: a window to
``virtual'' SUSY?}, preprint UAB-FT-397 [hep-ph/9607485].
\bibitem{QCDHaber}
Y. Grossman, H.E. Haber, Y. Nir,\, {\it Phys. Lett.}\, {\bf B 357} (1995) 630.
\bibitem{ALEPHL3}
D. Buskulick et al. (ALEPH Collab.)\, {\it Phys. Lett.}
\, {\bf B 343} (1995) 444; M. Acciarri et al. (L$3$ Collab.)\,
{\it Phys. Lett.}\, {\bf B 332} (1994) 201.
\bibitem{world}
L. Montanet et al. (Particle Data Group)\, {\it Phys. Rev.}
\, {\bf D 50} (1994) 355.
\bibitem{Neubert}
M. Neubert,\, {\it Phys. Rep.}\, {\bf  245} (1994) 259;
B. Grinstein, Lectures given at TASI 94, Boulder, CO, 29 May - 24 Jun 1994
[hep-ph/9411275].
\bibitem{Falk}
A.F. Falk, Z. Ligeti, M. Neubert, Y. Nir,\, {\it Phys. Lett.}\,
{\bf B 326} (1994) 145; L. Koyrakh,\, {\it Phys. Rev.},\, {\bf D 49}
 (1994) 3379 ; S. Balk, J.F. K\"orner, D. Pirjol,
K. Schilcher,\, {\it Z. Phys.}\, {\bf C 64} (1994) 37.
\bibitem{Czarnecki2}
A. Czarnecki, M. Jezabek, J. K\"uhn\, {\it Phys. Lett.}\,
{\bf B 346} (1995) 335; M. Jezabek, L. Motyka, preprint
UJ-TPJU 18/96 [hep-ph/9609352].
\bibitem{Grossman}
Y. Grossman, Z. Ligeti\, {\it Phys. Lett.}\, {\bf B 332} (1994) 373.
\bibitem{BSH}
M. B\"ohm, H. Spiesberger, W. Hollik,\, {\it Fortschr. Phys.}
{\bf 34} (1986) 687.
\bibitem{DHJJS}
A. Dabelstein, W. Hollik, R.A. Jim\'enez, C. J\"unger,
J. Sol\`a,\, {\it Nucl. Phys.}\, {\bf B 456} (1995) 75.
\bibitem{CD}
A. Czarnecki, S. Davidson,\, {\it Phys. Rev.} {\bf D 48} (1993) 4183.
\bibitem{Ricard}
R.A. Jim\'enez, J. Sol\`a, {\it Phys. Lett.}\, {\bf B 389} (1996) 53.
\bibitem{GJS}
J. Guasch, R.A. Jim\'enez, J. Sol\`a,\,{\it Phys. Lett.}
\,{\bf B 360} (1995) 47.
\bibitem{Anna}
ALEPH Collab., paper contributed to the ICHEP, Warsaw, Poland, 25-31 July 1996;
A. Pascual, {\it Updated Measurements of $b\rightarrow\tau\nu_{\tau}X$ decays},
to appear, in: Proc. of the 2nd Internat. Conf. on Hyperons, Montreal,
Qu\'ebec,
27-30 August 1996;
See also A. Pascual, {\it PhD Thesis}, Univ. Aut\`onoma de Barcelona (1995).
\bibitem{LEPEWW}
The Heavy Flavour sub-group of the LEP Electroweak
Working Group, preprint LEPHF/96-01, July 26, 1996.
\bibitem{RMiquel}
R. Miquel, talk given at the Particle Physics Seminar, CERN, October 8th, 1996.
\bibitem{SO10}
L.J. Hall, R. Rattazzi, U. Sarid,\, {\it Phys. Rev.} {\bf D 50}
 (1994) 7048; M. Carena, S. Pokorski,
C.E.M. Wagner,\,{\it Nucl. Phys.} {\bf B 426} (1994) 269;
R. Rattazzi, U. Sarid,\,
{\it Phys. Rev.} {\bf D 53} (1996) 1553.
\bibitem{Referee}
We thank our anonymous Referee for an insightful remark in this point.
\bibitem{LEPgluino}
The ALEPH Collab., preprint CERN-PPE/97-002 (January, 1997).
\bibitem{ConwayRoy}
J. Conway, talk given at SUSY 96, Univ. of Maryland,
College Park, USA, May 29th-June 1st 1996; M. Guchait, D.P. Roy\,\,
preprint TIFR/TH/96-58 [hep-ph/9610514].
\bibitem{CLEO}
M.S. Alam et al. (CLEO Collab.)\, {\it Phys. Rev. Lett.}\,{\bf 74} (1995) 2885.
\end{thebibliography}
\end{document}